\documentclass[%
 reprint,
 amsmath,amssymb,
 aps,
prb,
floatfix,
]{revtex4-2}

\usepackage{graphicx}
\usepackage{dcolumn}
\usepackage{amsmath}
\usepackage{multirow}
\usepackage{bm}
\usepackage{array}
\usepackage{siunitx}
\usepackage[utf8]{inputenc}
\usepackage{xcolor,soul}
\definecolor{bluehl}{RGB}{200,230,255} 
\sethlcolor{bluehl} 
\usepackage{url}
\usepackage{booktabs} 
\usepackage{textcomp}
\usepackage{afterpage}
\begin{document}


\title{Probing the Chirality of Trigonal Selenium and Tellurium by \\Spin and Orbital Hall Effects}
\author{Yuting Xiong}%
\author{Yingjie Hu}
\author{Wei Ren}
 \email{renwei@shu.edu.cn}
\author{Heng Gao}
\email{gaoheng@shu.edu.cn}
\affiliation{
 Physics Department, Material Genome Institute, Institute for Quantum Science and Technology,\\
International Centre of Quantum and Molecular Structures, Shanghai University, Shanghai 200444, China
}

\date{\today}

\begin{abstract}

{Chiral crystals exhibit enantiomer-dependent transport phenomena that generate pure spin or orbital currents, while the handedness sensitivity of spin and orbital Hall conductivities (SHC/OHC) remains insufficiently understood. Using first-principles calculations, we demonstrate that trigonal selenium and tellurium---prototypical chiral semiconductors--- exhibit opposite signs of the SHC/OHC tensor elements $\sigma_{yx}^{S_y}$ and $\sigma_{yx}^{L_y}$ between their left- and right-handed enantiomers. This behavior originates from the mirror operation relating the two structures, described by space groups $P3_221$ (left-handed) and $P3_121$ (right-handed). Although both enantiomers share identical band structures and four nonzero SHC/OHC tensor components, $\sigma_{yx}^{S_y}$ and $\sigma_{yx}^{L_y}$ reverse sign due to the antisymmetric transformation of the spin/orbital Berry curvature under the $M_{xy}$ mirror operation. More generally, for mirror-related enantiomorphic structures, selected SHC/OHC tensor components can exhibit symmetry-governed sign reversal.
For trigonal Se and Te, the calculated signs of these components can be directly correlated with the left- and right-handed structures under the chosen coordinate convention. These results clarify the symmetry origin of handedness-dependent SHC/OHC and suggest a possible route for correlating measurable SHC/OHC signals with structural handedness in specific chiral materials.
}

\end{abstract}


\maketitle

\section{Introduction}
Chirality plays a crucial role across physics, chemistry, and biology, as it describes entities that can exist in two non-superimposable forms known as enantiomers, where each form is a mirror image of the other \cite{Rikken2023,Siegel1998}. In 1894, Lord Kelvin characterized the property of chirality by asserting that a geometrical figure or object is considered chiral if its reflection in an ideal plane mirror cannot be superimposed onto the original object \cite{Kelvin1894, thomson2010baltimore}. Based on the symmetry requirements for chiral crystals, a subset of 65 space groups, known as the Sohncke groups, has been identified from the 230 crystal space groups \cite{fecher2022chirality}. In recent years, several fascinating aspects of chiral materials have been uncovered, such as the interaction between chirality and electronic topological states \cite{sanchez2019topological,schroter2019chiral,yao2020observation}, phonons \cite{cui2023chirality,zhang2023weyl,ohe2024chirality} and ferroelectricity \cite{long2020chiral,zeng2022axial,fu2022multiaxial}.

It is worth noting that not only has structural chirality itself been widely studied \cite{bousquet2025structural}, but the influence of chiral structures on optoelectronic \cite{fecher2022chirality,li2024large,duim2021chiral,chen2020probing,ding2024intense} and transport properties has also attracted significant attention. A representative optical manifestation is circular dichroism (CD), defined as the differential absorption of left- and right-circularly polarized light, which has been widely confirmed experimentally and computationally as an effective probe of structural handedness \cite{fecher2022chirality,li2024large,duim2021chiral}. Chen \textit{et al.} showed that chiral Te produces opposite CD signals for left- and right-handed crystals under circularly polarized light in high-harmonic spectra \cite{chen2020probing}. The significant CD effect has also been detected in nanoscale chiral AgBiS$_2$ nanocrystals with inorganic ligands \cite{ding2024intense}. In addition, a growing number of experimental and theoretical studies \cite{su2023chirality} suggest that left- and right-handed chiral structures exhibit opposite transport responses in a wide range of phenomena. The transport properties of chiral structures include spin \cite{sun2024colossal,yang2025chirality,xu2023chiral,hsieh2022helicity}, orbital \cite{yang2025chirality,adhikari2023interplay,gobel2025chirality}, electrical \cite{niu2023tunable,suarez2024odd,joseph2024chirality}, and thermal \cite{hsieh2022helicity,ramirez2024spin}. Experimentally, Niu \textit{et al.} measured the nonlinear electrical response of left- and right-handed 2D Te, including nonreciprocal electrical transport in the longitudinal direction and the nonlinear planar Hall effect in the transverse direction, and concluded that the nonlinear electrical response has opposite signs in left- and right-handed 2D Te \cite{niu2023tunable}. Moreover, Manuel \textit{et al.} showed that the non-linear susceptibility changes sign under spatial inversion, in agreement with its inversion-symmetry-odd nature \cite{suarez2024odd}. Recent first-principles calculations further showed that the Berry-curvature-dipole-induced nonlinear Hall effect in chiral materials (NbSi$_2$, Te, HgS, and topological multiple semimetal CoSi) exhibits opposite signs between enantiomers \cite{joseph2024chirality}.
Exploration of spin transport phenomena plays a crucial role in controlling the spin degree of freedom. In this context, spin transport in chiral materials has attracted considerable attention, with two representative mechanisms being the chirality-induced spin selectivity effect (CISS) \cite{dednam2023group,guo2012spin} and the spin Hall effect (SHE). Gupta \textit{et al.} performed first-principles calculations on a series of chiral crystals, including the intermetallic transition-metal disilicides $TM$Si$_2$ ($TM$ = V, Nb and Ta), the compounds $TMX$ ($TM$ = Co, Rh, Pd and Pt; $X$ = Al, Si and Ga), and structures containing Se and Te \cite{gupta2024current}, and demonstrated that electrons traversing these crystals exhibit the CISS effect \cite{yang2025chirality}, namely, spin polarization aligned with the current direction. Experimentally, Inui \textit{et al.} detected the CISS effect in the structurally chiral magnetic crystal CrNb$_3$S$_6$ by employing the inverse SHE as a probe \cite{inui2020chirality}. In contrast, the SHE generates spin accumulation at the lateral edges of a material, where the spin polarization is oriented perpendicular to the current, representing a transverse spin response mediated by spin-orbit coupling (SOC) \cite{zhao2022large}. In structurally chiral fermion semimetals $XY$ ($X$ = Co, Rh; $Y$ = Si, Ge), calculations of the spin Hall conductivity (SHC) have revealed that the structural handedness is correlated with the fermion chirality \cite{hsieh2022helicity}. Large SHC responses have also been reported in structurally chiral antiferromagnets Mn$_3$Sn and Mn$_3$Ge \cite{zhang2017strong}. Notably, two-dimensional structurally chiral monolayers of $MX$Te$_4$ ($M$ = metal; $X$ = S, Se, or Te) exhibit SHC-driven spin transport in which an in-plane electric field induces out-of-plane spin polarization \cite{zhao2023cluster}. 

A recent spin- and angle-resolved photoemission spectroscopy study by Sakano \textit{et al.} \cite{sakano2020radial} revealed a radial spin texture in Te dictated by its chiral crystal structure, with inward (outward) spin orientation in the right- (left-) handed enantiomers. Trigonal Se and Te are particularly suitable platforms for the present study because their chiral crystal structures are already known to host nontrivial electronic and spin textures. First-principles calculations have shown that trigonal Se and Te possess Weyl nodes near the Fermi level together with hedgehog-like spin textures around the H-point region \cite{hirayama2015weyl}. In addition, first-principles studies of trigonal Te have demonstrated handedness-dependent gyrotropic responses originating from Berry-curvature-related electronic-structure effects \cite{tsirkin2018gyrotropic}. The reversal of spin polarization with structural handedness provides a microscopic link between chirality and SOC, motivating investigations of how handedness influences SHC in group-VI chiral semiconductors such as Se and Te. Alongside the SHE, the orbital Hall effect (OHE) describes a transverse flow of orbital angular momentum ($L$) induced by an electric field \cite{go2018intrinsic,tanaka2008intrinsic}. Motivated by the enantiomer-dependent charge transport observed in chiral crystals, we aim to explore the interplay between structural handedness and spin/orbital transport phenomena and clarify the correlation between selected SHC/OHC tensor components and enantiomeric handedness in trigonal Se and Te.

In this work, based on first-principles calculations, we investigate the SHC and OHC of the inorganic chiral semiconductors Se and Te. By analyzing the symmetry-allowed tensor components and comparing them between left- and right-handed structures, we identify a distinct handedness-dependent response: an electric field applied along the $x$ direction induces spin or orbital currents along the $y$ direction, where the spin (or orbital) polarization is also along the $y$ axis but with opposite signs for opposite chiral crystals. Our results indicate that SHC and OHC provide a transport signature correlated with structural handedness in trigonal Se and Te. These results highlight the role of symmetry in governing spin and orbital Hall responses in chiral crystals and may motivate future experimental studies exploring chiral materials as potential platforms for SHE and OHE.

\section{ Computational methods}
First-principles calculations were performed using density functional theory (DFT) as implemented in the Vienna ab initio simulation package (VASP) \cite{kresse1993ab,kresse1996efficient}. The exchange-correlation functional was described using the generalized gradient approximation (GGA) in the Perdew-Burke-Ernzerhof (PBE) form \cite{perdew1996generalized}.  SOC was included in all calculations. A plane-wave cutoff energy of 500 eV was employed. The Brillouin zone was sampled using an $11\times11\times11$ $k$-point mesh for the PBE calculations and a $9\times9\times9$ mesh for the HSE06 calculations \cite{krukau2006influence}. The Bloch states obtained from the HSE06 calculations with SOC were projected onto maximally localized Wannier functions using the WANNIER90 package \cite{marzari2012maximally}. The SHC and OHC were calculated using the Kubo formula implemented in WANNIERTOOLS \cite{wu2018wanniertools}. Notably, the OHC was computed by modifying the WannierTools code to replace $S$ with $L$ in the routine for the Hall response. Further details are provided in the Supplemental Material, and this approach is supported by prior studies \cite{hu2024ferroelectric}. For the intrinsic SHC and OHC, a dense $100\times100\times100$ $k$-point mesh was used for the Brillouin-zone integration. The Kubo equation for SHC and OHC reads\cite{qiao2018calculation,sinova2015spin,roy2022unconventional,qu2023intrinsic}
\\\begin{equation}
\sigma_{i j}^{X_k}=e \hbar \sum_{n, k} f_{n, k} \sum_{m \neq n} \frac{2 \operatorname{Im}\left[\langle n\boldsymbol{k}| \hat{j}_i^{X_k}|m \boldsymbol{k}\rangle\langle m \boldsymbol{k}|\nu_j|n \boldsymbol{k}\rangle\right]}{\left(E_{n \boldsymbol{k}}-E_{m \boldsymbol{k}}\right)^2}
\end{equation}
where $e$ is the electron charge, $\hbar$ is the reduced Planck constant, $m$ and $n$ are band indexes, $E_{n k}$ and $E_{mk}$ are the eigenvalues, $f_{n,k}$ is the Fermi-Dirac distribution function and $\sigma_{i j}^{X_k}$ are spin ($X_{k}$=$S_{k}$) or orbital ($X_{k}$=$L_{k}$) Hall conductivity. Here, $j_{i}^{X_k}=\{S_{k},\nu_{i}\}/2$ is spin ($X_{k}$=$S_{k}$) or orbital ($X_{k}$=$L_{k}$) current operator, with the velocity operator defined as $\nu_j = (\partial_{k_j}\! H)\, / \,\hbar$. In the notation $\sigma_{i j}^{X_k}$, the subscripts $i$, $j$, $k$  denote Cartesian coordinates $x$, $y$, $z$, representing respectively the spin/orbit current propagation axis ($i$), the charge current flow direction ($j$) and the spin/orbit polarization orientation ($k$). The k-resolved term $\Omega_{i j}^{X_k}(\boldsymbol{k})$ is\\\begin{equation}
\Omega_{i j}^{X_k}(\boldsymbol{k})=\sum_n f_{n, k} \Omega_{n, i j}^{X_k}(\boldsymbol{k})
\end{equation}

and $\Omega_{n,i j}^{X_k}(\boldsymbol{k})$ is spin or orbital Berry curvature of band n term given by \\\begin{equation}
\Omega_{n, i j}^{X_k}(\boldsymbol{k})=\hbar^2 \sum_{m \neq n} \frac{2 \operatorname{Im}\left[\langle n \boldsymbol{k}| \hat{j}_i^{X_k}|m \boldsymbol{k}\rangle\langle m \boldsymbol{k}| \nu_{j}|n \boldsymbol{k}\rangle\right]}{\left(E_{n \boldsymbol{k}}-E_{m \boldsymbol{k}}\right)^2}.
\end{equation}

Owing to its formal similarity to the Kubo-like formula of the conventional Berry curvature, this term is often termed the spin Berry curvature (SBC) and orbital Berry curvature (OBC)  \cite{gradhand2012first}. The simplified Kubo equation is\\\begin{equation}
\sigma_{i j}^{X_k}=\frac{e}{\hbar} \int \frac{d k^3}{(2 \pi)^3} \Omega_{i j}^{X_k}(\boldsymbol{k})
.\end{equation}

\section{Results and Discussions}
\subsection{Chiral crystal structures of Se/Te
}

The trigonal unitcell of Se and Te comprises three atoms, each covalently bonded to two nearest neighbors, forming an inﬁnite helical screw along the $c$-axis as shown in Fig. 1. These helical screws align parallel to the $c$-direction and interact with adjacent helical screws via van der Waals interactions, constituting the bulk crystal. The handedness of helical screws dictates the space group: left-handed screws [(Fig. 1(a)] adopt space group $P3_221$ (No. 154), whereas right-handed screws [Fig. 1(b)] belong to $P3_121$ (No. 152) \cite{asendorf1957space}. The 3D helical arrows indicate whether the atomic screw extends along the $z$-direction in a clockwise (left-handed) or counterclockwise (right-handed) sense, consistent with the handedness illustrated by the hand icons. The handedness of trigonal Se/Te originates from their helical $3_2$/$3_1$ screw structures. The two enantiomers are related by the handedness-reversing $M_{xy}$ mirror mapping, which mathematically connects them but is not a symmetry operation of the chiral crystal itself. 

Although Se and Te crystallize in the same enantiomorphic space groups, their structural parameters differ significantly due to atomic size effects. The detailed structural parameters including intra-screw bond length ($r$),  inter-screw distance ($R$), and bonding angles ($\theta$) of Se and Te, are illustrated in Table 1. Our calculated structural parameters are consistent with experimental and previous theoretical results \cite{cheng2019large,keller1977effect,mccann1972compressibility,teuchert1975lattice}. Notably, the ratio of $r/R$ = 0.707 (Se)\( < \)0.841 (Te) (intra-screw to inter-screw distance) in Te is bigger than that in Se, indicating weaker structural anisotropy of tellurium \cite{cheng2019large}. While PBE-calculated lattice constants deviate by less than 3.3\% ($a$-axis) and 1.9\% ($c$-axis) from experimental results (Table I), these discrepancies are characteristic of generalized gradient approximations and do not impact our central conclusions. All electronic structure calculations employ experimentally refined lattice constants to ensure physical fidelity. 

\begin{figure}[ht]
    \centering
    \includegraphics[width=1\linewidth]{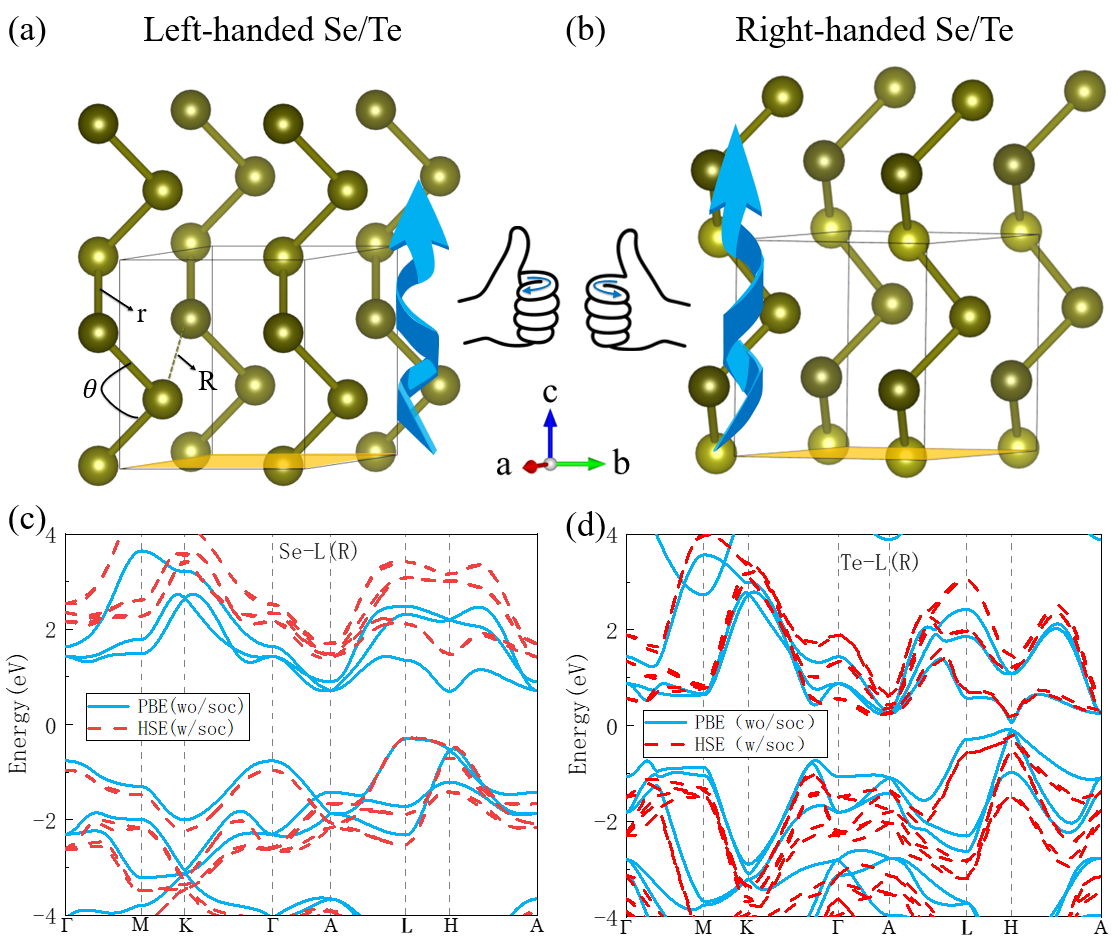}
    \caption{Crystal structures of chiral Se/Te. (a) and (b) show the schematic structures of left- and right-handed Se/Te, respectively. $\theta$, $r$, and $R$ in (a) denote the bond angle, intra-screw distance, and inter-screw distance, respectively. The 3D helical arrows that explicitly indicate
the screw sense (clockwise/counterclockwise) mirror operation $M_{xy}$ (yellow plane), respectively. The band structures of Se (c) and Te (d). HSE06 including SOC (red); PBE without SOC (cyan).}
    \label{figure:enter-label}
\end{figure} 

\begin{table}[ht]
    \centering
    \caption{Calculated (this work), experimental (Exp.), and other calculated of PBE lattice constants $a$ and $c$, the distance of intra-screw $r$ and inter-screw $R$, bond angle $\theta$, and bandgap. Values outside (inside) parentheses correspond to band gaps obtained from HSE06 with SOC (PBE without SOC).}
    \label{tab:lattice_parameters}
    
    \begin{tabular}{llcccccc}
        \toprule
         &  & $a$ & $c$ & $\theta$ & $r$ & $R$ & Bandgap \\
        \midrule
         & This work & 4.37 & 4.95 & 101.88 & 2.41 & 3.41 & 1.67 (0.99) \\
        Se & PBE \cite{cheng2019large} & 4.51 & 5.05 & 103.60 & 2.40 & 3.57 & 1.00 \\
         & Exp. \cite{mccann1972compressibility,teuchert1975lattice} & 4.37 & 4.95 & 103.07 & 2.37 & 3.44 & 2.00 \\
        \midrule
         & This work & 4.45 & 5.93 & 101.38 & 2.90 & 3.44 & 0.33 (0.15) \\
        Te & PBE \cite{cheng2019large} & 4.44 & 5.97 & 101.79 & 2.90 & 3.44 & 0.11 \\
         & Exp. \cite{cheng2019large,anzin1977measurement} & 4.45 & 5.93 & 103.27 & 2.83 & 3.49 & 0.32 \\
        \bottomrule
    \end{tabular}
\end{table}

\subsection{Electronic structures of chiral Se/Te}
The band structures of trigonal Se and Te are presented in Fig.~1. Se exhibits a semiconducting character with a band gap of 1.67 eV in the HSE06 calculation including SOC (red curves), where the valence-band maximum is located at the L point, and the conduction-band minimum is near the A point. Te exhibits a narrow direct band gap of 0.33 eV at the H point. In addition, the third and fourth conduction bands cross at H, forming a Weyl-type crossing. Such band crossings are consistent with previous reports of symmetry-enforced Weyl points in Se and Te, which have been systematically analyzed and classified in earlier studies \cite{hirayama2015weyl,gatti2020radial,tsirkin2017composite}. To analyze the orbital character of the electronic states, we also calculated the orbital-projected band structures of Se and Te without SOC, as shown in Fig.~S1 \cite{Supplemental_2025}. The results indicate that the electronic states near the Fermi level in both Se and Te are predominantly derived from $p$ orbitals, while the $s$ orbitals contribute only to deep valence bands located below $-8$ eV.

For both Se and Te, the inclusion of SOC leads to distinct band splitting in their band structures, reflecting the combined effect of SOC and inversion-symmetry breaking in these chiral crystals. Owing to its stronger atomic SOC ($5p > 4p$), Te exhibits larger SOC-induced band splittings than Se. The calculated band gaps of Se and Te without SOC are $0.99$ eV and $0.15$ eV, respectively. Our results are in agreement with previous HSE06 results \cite{cheng2019large}, as shown in Table~I. For comparison, the band gaps calculated by PBE are significantly smaller than the experimental band gaps of Se ($2.0$ eV) and Te ($0.3230$ eV), owing to the well-known underestimation of band gaps by the PBE approximation. Therefore, to achieve better agreement with experiment, the following main-text analysis is based on the HSE06 results. Although the PBE functional is computationally efficient, it may affect the quantitative accuracy of the calculated electronic structure and the related response functions. Further details are provided in Sec.~II of the Supplemental Material \cite{Supplemental_2025}.

Motivated by these symmetry constraints, Se and Te emerge as prototypical chiral systems for investigating the interplay between structural handedness and transport phenomena. We show that selected SHC/OHC tensor components, originating from the corresponding spin/orbital Berry-curvature-related quantities, can be directly correlated with structural handedness in these systems. Importantly, these quantities exhibit handedness-dependent sign reversal between the two enantiomers, while the band dispersions remain invariant. This makes SHC/OHC a symmetry-governed transport signature of handedness in the present nonmagnetic chiral systems.

\subsection{Spin Hall conductivity of  Se/Te}
Se and Te are both nonmagnetic chiral crystals with trigonal symmetry, belonging to Laue group of $\bar{3}m$.  According to the symmetry rule of the non-magnetic Laue group \cite{roy2022unconventional}, the spin Hall conductivity tensor can be expressed as
\begin{equation}
\sigma^{S_x}=\left(\begin{array}{ccc}
\sigma_{x x}^{S_x} & 0 & 0 \\
0 & \sigma_{y y}^{S_x} & \sigma_{y z}^{S_x} \\
0 & \sigma_{z y}^{S_x} & 0
\end{array}\right) ,
\end{equation}
\begin{equation}
    \sigma^{S_y}=\left(\begin{array}{ccc}
0 & \sigma_{x y}^{S_y} & \sigma_{x z}^{S_y} \\
\sigma_{y x}^{S_y} & 0 & 0 \\
\sigma_{z x}^{S_y} & 0 & 0
\end{array}\right) ,
\end{equation}
\begin{equation}
\sigma^{S_z}=\left(\begin{array}{ccc}
0 & \sigma_{x y}^{S_z} & 0 \\
\sigma_{y x}^{S_z} & 0 & 0 \\
0 & 0 & 0
\end{array}\right).
\end{equation}
Only four independent non-zero SHC tensors exist due to trigonal symmetry \cite{Gallego2019}. It should be emphasized that non-zero tensor elements have the equivalence relations: $
\sigma_{xy}^{S_z} = -\sigma_{yz}^{S_x},\quad
\sigma_{yx}^{S_y} = \sigma_{yy}^{S_x} = \sigma_{xy}^{S_y} = -\sigma_{xx}^{S_x},\quad
\sigma_{zx}^{S_y} = -\sigma_{zy}^{S_x},\quad
\sigma_{xz}^{S_y} = -\sigma_{yz}^{S_x}.
$
\begin{figure}[ht]
    \centering
    \includegraphics[width=1\linewidth]{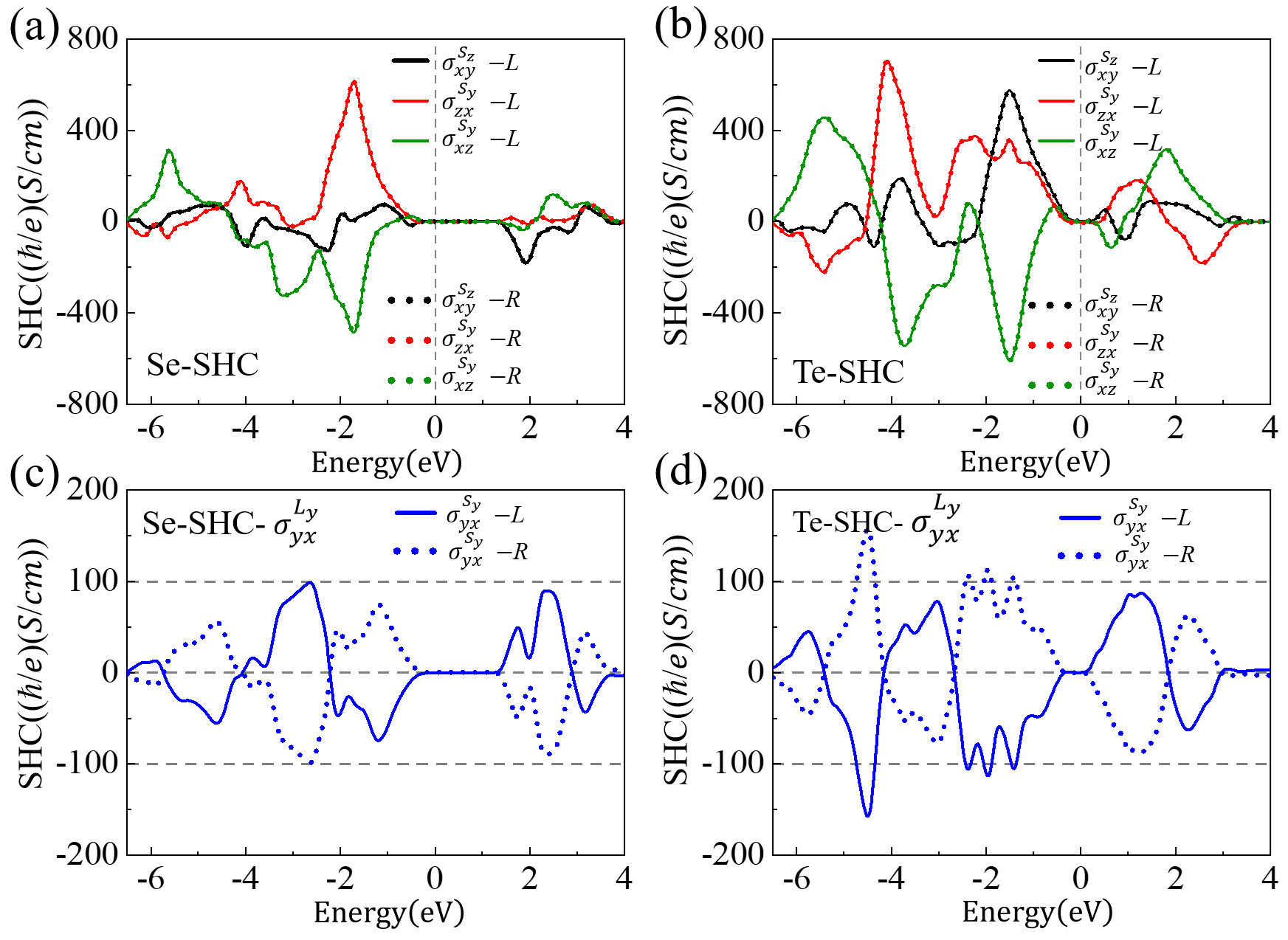}
    \caption{The non-zero independent SHC tensors $\sigma_{xy}^{S_z}$, $\sigma_{zx}^{S_y}$ and $\sigma_{xz}^{S_y}$ of left- and right-chiral trigonal (a) Se and Te (b). The comparison of $\sigma_{yx}^{S_y}$ components for left- and right-handed (c) Se and (d) Te. }
    \label{fig:placeholder}
\end{figure}

The independent non-zero SHC tensors of Se and Te with opposite handedness were calculated and shown in Figs 2(a)-(d). It turns out that the significant difference between four independent SHC tensors for the same material indicates strong anisotropy of SHC of Se and Te. Comparing the SHC between Se and Te with the same tensor component, the peak values of SHC in Te exceed those in Se by $\sim$$50$\% [e.g., $157.7$ $(\hbar/ e) (\mathrm{S} / \mathrm{cm})$ vs $98.6$ $(\hbar/ e) (\mathrm{S} / \mathrm{cm})$ for Se]. This is due to the stronger SOC of Te than Se. Furthermore, it is worth noting that in Figs 2(c) and (d), the blue lines show the variation of SHC with energy for Se and Te of the left- (solid line) and right- (dashed line) handedness, respectively. The SHC of Se left- and right-handed shows a distinct trend: a nonzero tensor element, denoted as $\sigma_{yx}^{S_y}$, exhibits opposite values in the opposite handedness. There is a positive value of $98.6$ $(\hbar/ e) (\mathrm{S} / \mathrm{cm})$ at the Fermi energy of $2.67$ eV for left-handed Se (solid line) and a negative value of $-98.6$ $(\hbar/ e) (\mathrm{S} / \mathrm{cm})$ at the same energy for right-handed Se (dashed line).

In the case of Te, the conclusion that the tensor element exhibits opposite signs arising from structural handedness switching remains valid. Figure 2(d) reveals that the unique tensor element $\sigma_{yx}^{S_y}$ again shows opposite signs for left- and right-handed Te,  even though its largest peak appears near $-4.5$ eV, with an increased magnitude of $157.7$ $(\hbar/ e) (\mathrm{S} / \mathrm{cm})$ compared to Se. Additionally, semiconducting Se displays a zero-SHC plateau near $E_{F}$ (in the width of band gap). 
\begin{figure}[ht]
    \centering
    \includegraphics[width=1\linewidth]{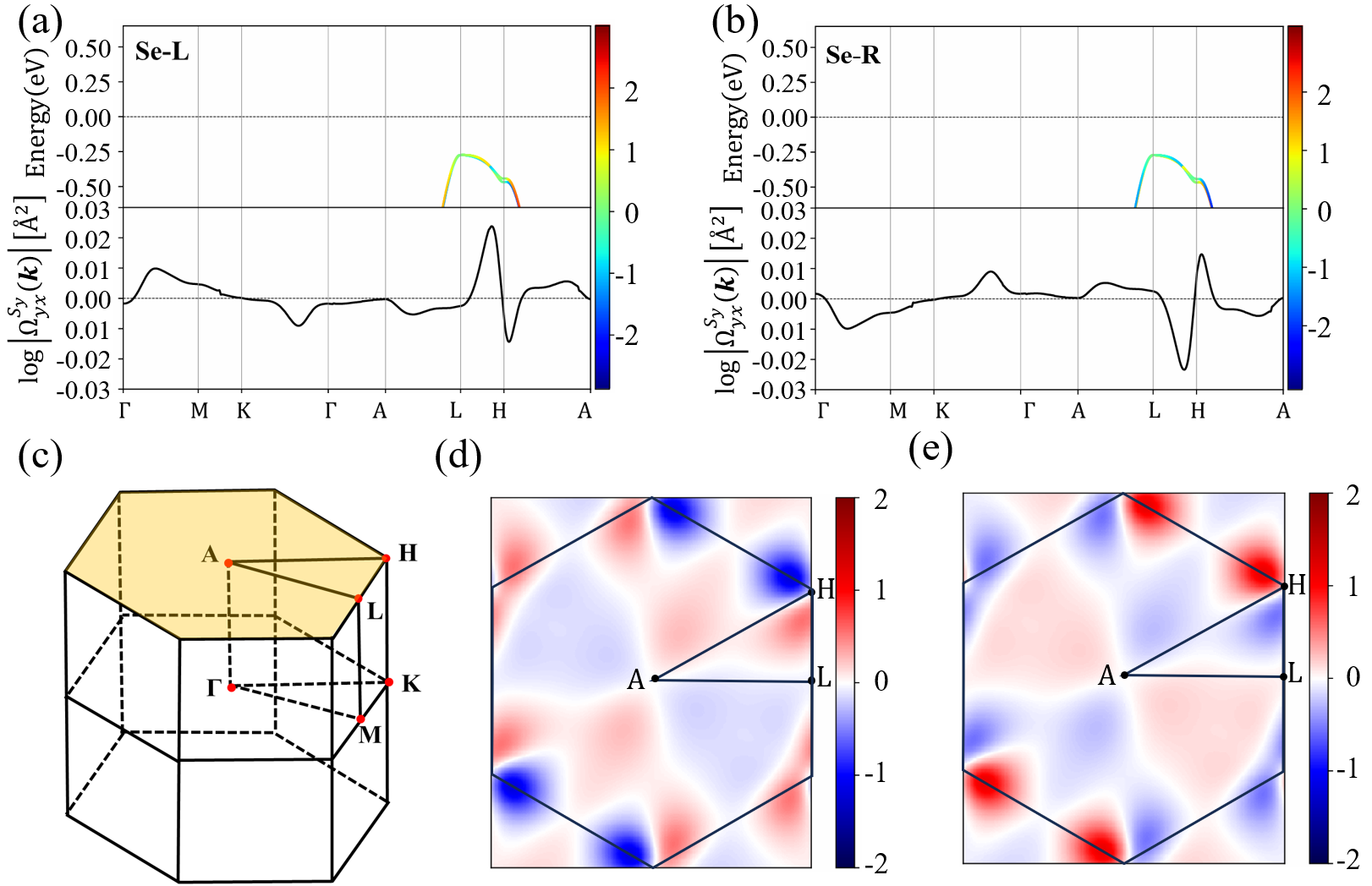}
    \caption{The band-projected spin Berry curvature (top) $\Omega_{n,yx}^{S_y}(k)$ on a log scale, $k$-resolved spin Berry curvature (bottom) $\Omega_{yx}^{S_y}(k)$ in (a) left- and (b) right-handed Se. The $\Omega_{yx}^{S_y}(k)$ at the Fermi level on a log scale in a slice of $k_z = 0.5 \times ({2\pi}/{c})$ for (c) left- and (d) right-handed Te.}
    \label{fig:placeholder1}
\end{figure}
To further analyze and illustrate the handedness dependence of the SHC tensor element $\sigma_{yx}^{S_y}$ in Se and Te structures, we calculated $\Omega_{n,yx}^{S_y}(k)$ and $\Omega_{yx}^{S_y}(k)$ for left- and right-handed Se structures, as shown in Figs~3(a),  (b), and~(d), (e), respectively. The yellow surface in the illustration represents the cross-section of the reciprocal space passing through point H in Fig 3(c). The numerical comparison of $\Omega_{n,yx}^{S_y}(k)$ and $\Omega_{yx}^{S_y}(k)$ on the high symmetry path of the left- and right-handed Se structures is done, which proves that the non-zero SHC tensor element $\sigma_{yx}^{S_y}$ in the left- and right-handed Se structures is opposite due to the opposite $\Omega_{yx}^{S_y}$ of the left- and right-handed Se. The $\Omega_{yx}^{S_y}(k)$ of Se shows a peak near the H point. Therefore, analyzing the distribution of $\Omega_{yx}^{S_y}(k)$ on the slice plane through the H point at the Fermi level, we found that the $\Omega_{yx}^{S_y}$ distribution of left- and right-Se in reciprocal space is completely opposite. As shown in Figs 3(d) and (e), the left- and right-handed Se structures are recognized by their colors, with the logarithmic (base 10) values of $\Omega_{yx}^{S_y}(k)$. 

\begin{figure}[ht]
    \centering
    \includegraphics[width=1\linewidth]{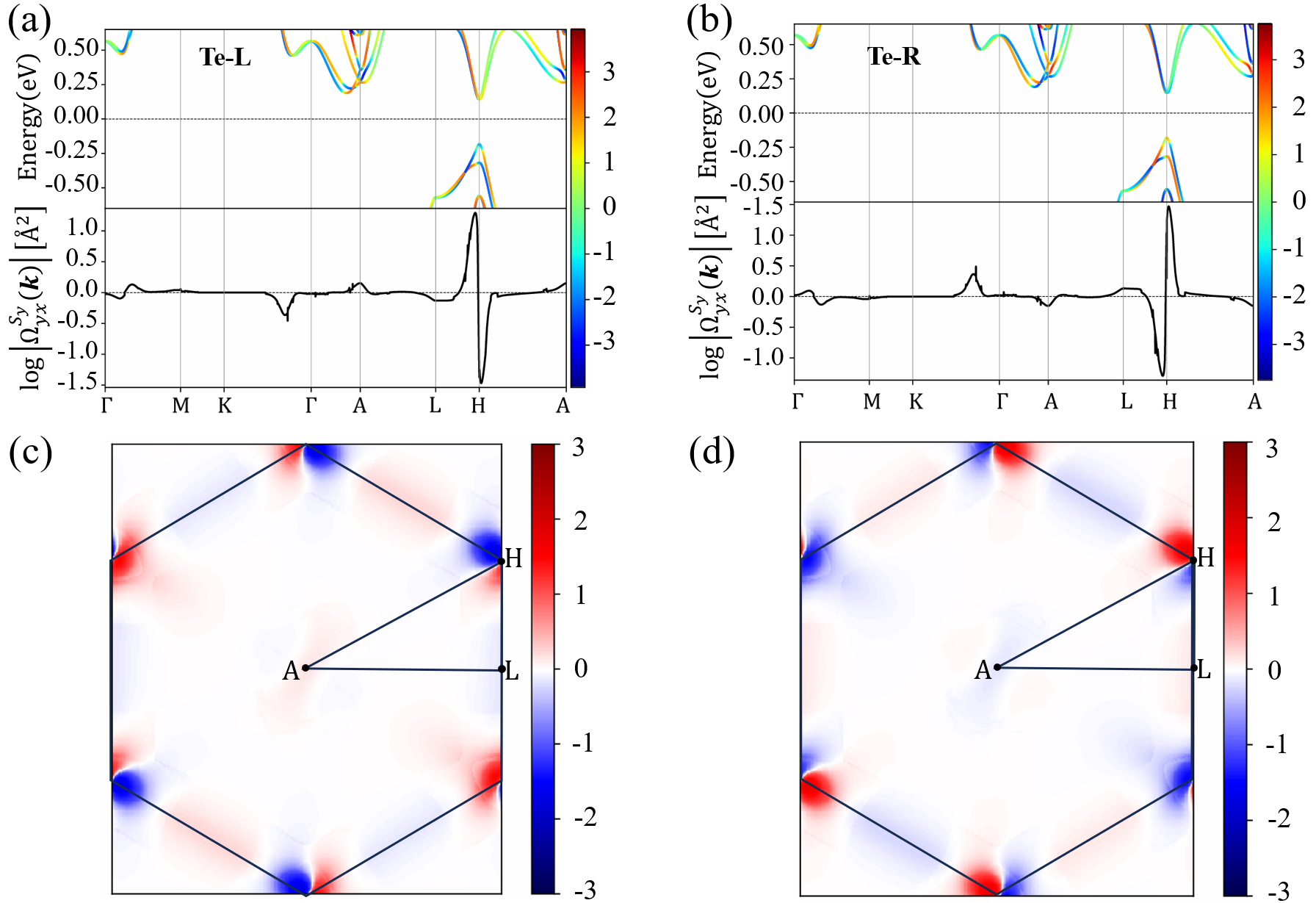}
    \caption{The band-projected spin Berry curvature (top) $\Omega_{n,yx}^{S_y}(k)$ on a log scale, $k$-resolved spin Berry curvature (bottom) $\Omega_{yx}^{S_y}(k)$ in (a) left- and (b) right-handed Se. The slice plane $k_z = 0.5 \times (2\pi/c)$ is displayed with yellow color in (c). The $\Omega_{yx}^{S_y}(k)$ at the Fermi level on a log scale in a slice of $k_z = 0.5 \times (2\pi/c)$ for (d) left- and (e) right-handed Se.}
    \label{fig:placeholder1}
\end{figure}
From Eqs. (2) and (3), it is theoretically predicted that the {$\Omega_{yx}^{S_y}(k)$ and $\Omega_{n,yx}^{S_y}(k)$ of $\sigma_{yx}^{S_y}$ tensor element for the two enantiomers, should also exhibit opposite sign. This is reflected in Figs. 4(a) and (b) for Te. Figures 4(c) and (d) depict $k$-slice plots of the first Brillouin zone of Te at reciprocal space $ k_z = 0.5 \times ({2\pi}/{c}) $, where the opposite sign of $\Omega_{yx}^{S_y}(k)$ manifests as opposite distributions of the curvature. To provide a detailed analysis, we focus on the $\Omega_{yx}^{S_y}(k)$ of left-handed Te [Figure 4(a)] and its corresponding $\Omega_{yx}^{S_y}(k)$} [Figure 4(c)]. At point A, $\Omega_{yx}^{S_y}(k)$ is relatively small, whereas along the path from A to H ([solid line in Fig.~4(a)], its magnitude increases significantly and becomes negative near H. This behavior is represented by the blue-gradient region around the H point in Fig.~4(c).

Overall, our first-principles calculations show that the selected SHC tensor component $\sigma_{yx}^{S_y}$ in chiral Se and Te exhibits opposite signs for the two enantiomers. Its magnitude and sign are governed by the underlying distribution of $\Omega_{yx}^{S_y}(k)$, revealing the symmetry origin of the handedness-dependent SHC response in these materials. This indicates that the handedness of the chiral crystal has a profound effect on spin-current generation in these materials. In other words, SHC is governed by the SBC, whose symmetry transformation properties differ from those of the Berry curvature relevant to the anomalous Hall effect. In nonmagnetic chiral systems, the SBC is invariant under spatial inversion but may change sign under mirror operations, depending on the tensor component. Consequently, for the mirror-related enantiomorphic structures considered here, the tensor component $\sigma_{yx}^{S_y}$ exhibits opposite signs for the left- and right-handed structures. This behavior is consistent with previous studies of trigonal Te showing that handedness can govern gyrotropic responses through Berry-curvature-related electronic-structure effects \cite{tsirkin2018gyrotropic}.

\subsection{Orbital Hall conductivity of Se/Te  }
We note that the symmetry constraints derived for the SHC tensor carry over directly to the OHC tensor upon replacing the $S$ by the $L$, since both transform as axial vectors under point-group operations. The mechanistic difference (OHC can exist without SOC, while SHC typically requires it) affects whether a nonzero response is realized in a given material, but does not change the symmetry-imposed form of the conductivity tensors \cite{go2018intrinsic,tanaka2008intrinsic}. To calculate the OHC using the Kubo formula [Eqs. (1) and (3)], the orbital current operator is $J_i^{L_k} = (L_k \nu_i + \nu_i L_k)/2$ where $L_k$ denotes the $k$-component of the orbital angular momentum. The computational methodology parallels that of the spin Hall conductivity, with the spin Pauli matrices substituted by $L$ matrices. Analogous to SHC, the non-zero tensor components of OHC in Se and Te exhibit patterns similar to those of the SHC. 

\begin{figure}[ht]
    \centering
    \includegraphics[width=1\linewidth]{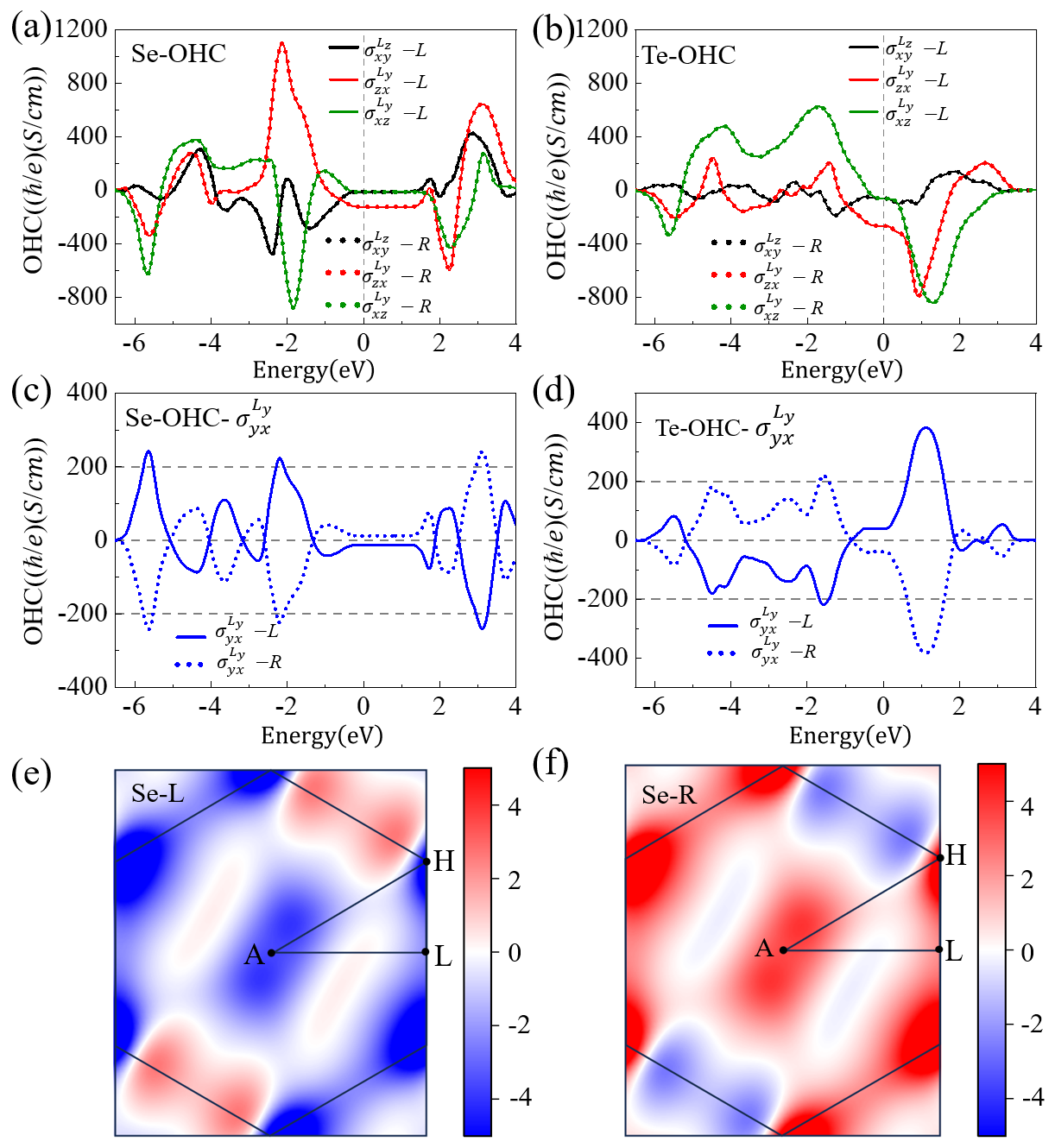}
    \caption{The OHC of (a, c) Se and (b, d) Te. Four non-zero independent tensor components ($\sigma_{xy}^{L_z}$, $\sigma_{yx}^{L_y}$, $\sigma_{zx}^{L_y}$, $\sigma_{xz}^{L_y}$) for left- (solid lines) and right- (dash line) handed are represented by black, blue, red, and green lines. The comparison of  $\sigma_{yx}^{L_y}$ for left- and right-handed (c) Se and (d) Te. The $k$-resolved OBC $\Omega_{yx}^{L_y}$  at Fermi level in a slice of $ k_z = 0.5 \times ({2\pi}/{c}) $ for (e) left- and (f) right-handed Se.}
    \label{fig:placeholder3}
\end{figure}

The OHC of chiral Se and Te are calculated, revealing four non-zero tensor components $\sigma_{xy}^{L_z}$, $\sigma_{yx}^{L_y}$, $\sigma_{zx}^{L_y}$, $\sigma_{xz}^{L_y}$ as functions of energy, as shown in Fig. 5. Solid and dashed lines represent left- and right-handed structures, respectively, with black, blue, red, and green curves corresponding to the four distinct non-zero OHC tensor components. Notably, in Se, the OHC tensor $\sigma_{zx}^{L_y}$ reaches a maximum value of $1099.2$ $(\hbar/ e) (\mathrm{S} / \mathrm{cm})$ at $-2.13$ eV, which is of the same order of magnitude as those observed in conventional semiconductors like Si and Ge \cite{baek2021negative}. Remarkably, the enantiomer-dependent behavior of the OHC $\sigma_{yx}^{L_y}$  persists in both chiral Se and Te, handedness-dependent sign reversal in the SHC, as demonstrated in Figs. 5(b) and (d). For Se, $\sigma_{yx}^{L_y}$ attains a value of $224.1$ $(\hbar/ e) (\mathrm{S} / \mathrm{cm})$ at $-2.19$ eV for the left-handed structure, while the right-handed structure exhibits a of $-224.1$ $(\hbar/ e) (\mathrm{S} / \mathrm{cm})$ at the same energy. Similarly, in Te, $\sigma_{yx}^{L_y}$ reaches $382.9$ $(\hbar/ e) (\mathrm{S} / \mathrm{cm})$ at $1.11$ eV for the left-handed structure, with the right-handed counterpart showing a corresponding minimum of $-382.9$ $(\hbar/ e) (\mathrm{S} / \mathrm{cm})$. Similar to the SHC, the OHC originates from the contributions of the orbital Berry curvature (OBC). Therefore, Figs. 5(e) and (f) show the distribution of the OBC in momentum space at the Fermi level on the $ k_z = 0.5 \times ({2\pi}/{c}) $ plane for the left- and right-handed Se structures, respectively. The OBC distributions along high-symmetry points and the cross-sectional distribution in Te, as well as their variation along high-symmetry paths, are presented in Fig. S8.

\subsection{Symmetry analysis of reversal SHC and OHC  with opposite handedness}
\begin{figure}[ht] 
    \centering
    \includegraphics[width=1\linewidth]{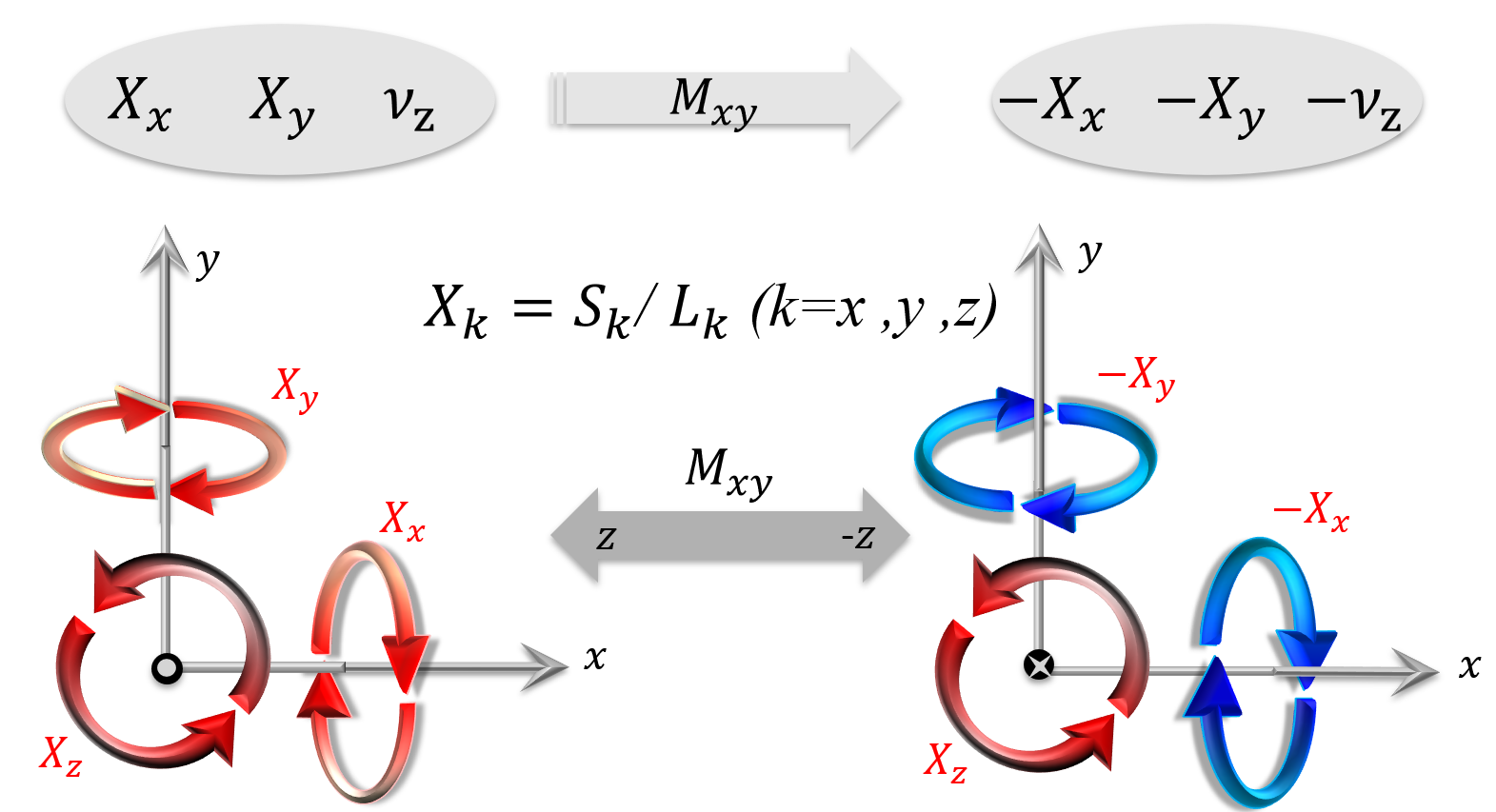}
    \caption{Schematics of the transformation relationship of $S_k$ and $L_k$ under mirror symmetry operation $M_{xy}$, where \textit{k} denotes the direction of S or L along the \textit{x}, \textit{y}, or \textit{z} axis. }
    \label{fig:placeholder4}
\end{figure}

The calculations of SHC and OHC for chiral Se and Te reveal a consistent result: the sign of the nonzero tensor component $\sigma_{yx}^{X_y}$ reverses between the two enantiomers. However, both SHC and OHC exhibit four distinct nonzero tensor elements, namely $\sigma_{xy}^{X_z}$, $\sigma_{yx}^{X_y}$, $\sigma_{zx}^{X_y}$, and $\sigma_{xz}^{X_y}$, where $X_y=S_y$ and $X_z=S_z$ for SHC, and $X_y=L_y$ and $X_z=L_z$ for OHC. Among these components, only $\sigma_{yx}^{X_y}$ exhibits a value handedness-dependent sign reversal between the left- and right-handed structures. Figure~6 illustrates the symmetry origin of this behavior. To analyze it, we consider the key quantities entering Eq.~(1): the spin/orbital current operator $\hat{j}_i^{X_k}=\{X_k,\nu_i\}/2$ and the velocity operator $\nu_j=(\partial_{k_j}H)/\hbar$. The sign reversal of $\sigma_{yx}^{X_y}$ is determined by the transformation of $X_y$ and the relevant velocity components under the handedness-reversing mirror mapping $M_{xy}$. Such symmetry constraints on response tensors are consistent with the general tensor analysis of symmetry-adapted quantities in nonmagnetic crystals \cite{Gallego2019}, while the use of Berry-curvature-related quantities in Hall-response calculations follows the standard Kubo formalism \cite{gradhand2012first}. As illustrated in Fig.~6, $X_y$ changes sign under $M_{xy}$, whereas $X_x$ and $X_z$ remain invariant. For the velocity operator, $\nu_x$ and $\nu_y$ remain unchanged, while $\nu_z$ changes sign. Consequently, for the tensor component $\sigma_{yx}^{X_y}$, the sign change of $X_y$, together with the invariance of $\nu_x$ and $\nu_y$, causes the current operator $\hat{j}_y^{X_y}=(X_y\nu_y+\nu_yX_y)/2$ to reverse sign between the two enantiomers. The variations of the other tensor components are listed in Table~S1. This explains why the handedness-dependent sign reversal appears specifically in $\sigma_{yx}^{X_y}$.

\section{Conclusion}
In summary, we investigated the SHC and OHC of enantiomeric Se and Te with opposite handedness. First-principles calculations demonstrate that chiral trigonal Te and Se exhibit handedness-dependent sign reversal in the SHC/OHC tensor element $\sigma_{yx}^{S_y}$/$\sigma_{yx}^{L_y}$. This phenomenon originates from the opposite SBC/OBC distributions in the two enantiomers, while other symmetry-allowed tensor components remain common to both enantiomorphs. Crucially, the sign reversal of $\sigma_{yx}^{S_y}$/$\sigma_{yx}^{L_y}$ leads to opposite transverse spin/orbital current polarizations for the two structures. 
For the specific Se/Te systems studied here, our results further show that, when the experimental axes and measurement directions are chosen consistently with the theoretical tensor definition, the sign of the measured SHC/OHC component can be directly correlated with the left- and right-handed structures. In this respect, our conclusion is also consistent with previous studies of trigonal Te showing handedness-dependent gyrotropic and spin-texture responses \cite{tsirkin2018gyrotropic,sakano2020radial}. While our analysis focuses on the enantiomorphic pair $P3_121$/$P3_221$ for Se/Te, the handedness-dependent sign reversal of SHC/OHC should be understood as a symmetry-governed consequence for selected tensor components of mirror-related enantiomorphic structures, rather than as a universal criterion for chirality. A group-theoretic analysis of operator transformations under the specific symmetries of each space group can further identify handedness-sensitive tensor components for the 11 enantiomorphic space-group pairs considered here, as summarized in Tables~S II. and S III. More generally, the present Se/Te results suggest a possible route for other enantiomorphic chiral space groups: namely, to identify, through symmetry analysis and material-specific calculations, the tensor components and measurement directions for which the sign of SHC/OHC can be correlated with structural handedness. In this sense, structural handedness may serve as a useful degree of freedom for tuning spin and orbital transport responses in specific chiral materials.

\begin{acknowledgments}

This work was supported by National Natural Science Foundation of China (Grants No. 12204299, No. 12074241, No. 11929401, No. 52130204, and No. 12274278).
\end{acknowledgments}

\section*{data availability}
The data that support the findings of this article are not publicly available. The data are available from the authors upon reasonable request.

\bibliography{bibfile.bib}

\end{document}